\documentclass[final,3p,times,twocolumn]{elsarticle}

 \usepackage{graphicx}

\usepackage{amssymb}
\usepackage{color}
\usepackage{listings}
\usepackage{mdwlist}

\definecolor{mygreen}{rgb}{0,0.6,0}
\definecolor{mygray}{rgb}{0.5,0.5,0.5}
\newtheorem{definition}{Definition}

\begin{document}

\begin{frontmatter}



\title{Component Matching Approach in Linking Business and Application Architectures}


\author{Suresh Kamath}

\address{MetLife Inc. \\ 300 Davidson Avenue, Somerset NJ 08873}

\begin{abstract}
The development of an IT strategy and ensuring that it is the best possible one for business is a key problem many organizations face. This problem is that of linking business architecture to IT architecture in general and application architecture specifically. In our earlier work we proposed Category theory as the formal language to unify the business and IT worlds with the ability to represent the concepts and relations between the two in a unified way.  We used rCOS as the underlying model for the specification of interfaces, contracts, and components.  The concept of pseudo-category was then utilized to represent the business and application architecture specifications and the relationships contained within. The linkages between them now can be established using the matching of the business component contracts with the application component contracts. However the matching was based on manual process and in this paper we extend the work by considering automated component matching process. The ground work for a tool to support the matching process is laid out in this paper.
\end{abstract}


\end{frontmatter}


\section{Introduction}
\label{sec:intro}
The development of an IT strategy and ensuring that it is the best possible one for business is a key problem many organizations face. This problem is that of linking business architecture to IT architecture in general and application architecture specifically. In our earlier work  \cite{Kamath:ICISA.2011} ,\cite{Kamath:WICSA.2011.12}, and \cite{IGI:978-1-4666-2199-2} we proposed Category theory as the formal language to unify the business and IT worlds with the ability to represent the concepts and relations between the two in a unified way.  We used rCOS \cite{report406} as the underlying model for the specification of interfaces, contracts, and components.  The concept of pseudo-category \cite{DBLP:journals/jcst/Lu05} was then utilized to represent the business and application architecture specifications and the relationships contained within. The linkages between them now can be established using the matching of the business component contracts with the application component contracts. However the matching was based on manual process and in this paper we extend the work by considering automated component matching process. In this paper we lay the ground work for a tool to support the matching process.

The rest of the paper is organized as follows. In Section \ref{sec:prob} we define the problem to be addressed followed by Section \ref{sec:related} wherein we review some of the related work in Component and Contract matching and software adaptability aspects. We discuss several use cases in Section \ref{sec:usecase} followed by the proposed framework in Section \ref{sec:framework}. We provide concluding remarks in Section  \ref{sec:conclusion}

\section{Background and Problem Definition} 
\label{sec:prob}
\subsection{Background}
We need some basic definitions that will be of use later to describe a business or application architecture. Following the rCOS approach  \cite{report406} leading up to the definition of a component, we have:

\begin{definition}
Interface I = (FDec, MDec), where FDec is a set of field or attribute declarations and MDec is a set of method declarations. An attribute {\bf a} has a type {\bf T} and a method {\bf m} can have input and output attributes, each of which will belong to a type {\bf U} and {\bf V}. 
\end{definition} 

We next proced to define a Contract that satisfies an interface I specified above:
\begin{definition}
A Contract is specified as Ctr = (I, Init, $\Phi$, Prot), where I is the interface satisfied by the contract, Init is an assignment of initial values to a set of attributes, if any. $\Phi$ is a function that maps each method of I to a specification, and Prot is an interaction protocol.
\end{definition} 
While the interface specifies only the syntax, a contract additionally specifies the functionality, what happens in a method. A protocol is nothing but a set of sequences of call events on the methods of the contract, e.g., $m_1 (x_1$) $m_2(x_2$)…$m_k(x_k$), where $m_i \in$ MDec. A general contract (GCtr) extends a contract Ctr with a set of private method declarations and their specifications.

Now we are ready to define a component:
\begin{definition}
A component (C) is an implementation of the interface of a general contract (called provided services).
This implementation may require methods (services) provided by other components and are called required services. Formally a component C = (GCtr,InMDec), where InMDec is the set of required methods by the component, also called required interfaces.
\end{definition} 

In order to define an architecture based on these elements we need to compose different components to provide the different functionality provided by the architecture. Interfaces, Contracts and Components that we have defined can be composed, we refer to \cite{report298} for details.

The basic idea in representing business architecture is to use the concept of interface described earlier to capture the basic services provided by a single business task or a business process. In this context it is also possible to extend an interface to add additional methods. The composition of interfaces will be used to represent the aggregate entity called the business capability.

Consider now a structure using M number of interface specifications represented as $I_i$, where  $1 <i < M$. An interface $I_i$ may extend ({\bf ext}) another interface $I_j$, that is the component $I_2$ has one or more attributes or methods different from $I_1$. An interface may be composed ({\bf cmp}) of two or more other interfaces, that is we can merge two interfaces to combine them. We have shown that such a structure forms a pseudo category, see \cite{IGI:978-1-4666-2199-2}. Please note that we need the {\bf ext} operation to support pre-existing interfaces in the business architecture, new interfaces would have been composed.

\begin{definition}
Business Architecture : is a collection of interfaces $I_i$, with $1 \leq i \leq M$, is a pseudo- category BA = (I, $M_I$, $G_I$, $T_I$), where:
\begin{itemize*}
\item I is a set containing  the interface containing the interface specifications $I_i$;
\item $M_I$ contains the morphisms f: $I_i$ $\stackrel{t}{\rightarrow}$ $I_j$, where t = \{ext, cmp\};
\item The set $G_I$ is \{ext, cmp\};
\end{itemize*}
\end{definition}
Given that $I_2 \thinspace  extends \thinspace  I_1$  (($I1 \stackrel{ext}{\longrightarrow}  I_2$) and $I_3 \thinspace  extends \thinspace  I_2$ ($I_2  \stackrel{ext}{\longrightarrow}  I_3$), we can see that $I_3 \thinspace  extends \thinspace   I_1$ ($I_3  \stackrel{ext}{\longrightarrow}  I_1$). The same applies for the cmp (composition) relation. If, for example, $I_2$ has $I_1$ in its composition and $I_3$ has $I_2$ in its composition, then $I_3$ also has $I_1$ in its composition.

We use the component specifications as the objects for representing the application architecture. The application architecture is a pseudo-category with the morphism types {\bf use}, {\bf cmp}. The {\bf use} stands for the chaining and the {\bf cmp} stands for the disjoint composition, see \cite{report298}. 
\begin{definition}\label{def:aa}
Application Architecture: This is a collection of the component specifications $C_i$, with $1\leq i \leq N$, is a pseudo-category AA = (C, $M_C$, $G_C$, $T_C$), where:
\begin{itemize*}
\item C is a set containing the component specifications $C_i$;
\item $M_C$ contains the morphisms g: $C_i \stackrel{t}{\longrightarrow} C_j$,  where t  =  \{use,cmp\};
\item The set $G_C$ is \{use,cmp\};
\item The set $T_C$ contains use x use $\rightarrow$  use and cmp x cmp $\rightarrow$  cmp
\end{itemize*}
\end{definition}
Given that $C_2 \thinspace uses \thinspace C_1$ ($C_1\stackrel{use}{\longrightarrow} C_2$) and $C_3 \thinspace uses \thinspace C_2$, ($C_2  \stackrel{use}{\longrightarrow} C_3$ ), we can see that $C_3 \thinspace uses \thinspace C_1$ ($C_3 \stackrel{use}{\longrightarrow}  C_1$ ). The same applies for the cmp (composition relation. If, for example, $C_2$ has $C_1$ in its composition and $C_3$ has $C_2$ in its composition, then $C_3$ also has $C_1$ in its composition.

Having defined how to represent the business and application architecture, we will now provide the mechanism to link the two. This will be using the concept of a pseudo-functor, which again depends on the typed functor between two categories.

\begin{definition}
A typed functor F between two typed categories $K_1$ = ($O_1$; $M_1$; $G_1$; $T_1)$ and $K_2$ = ($O_2$; $M_2$; $G_2$; $T_2)$ is defined as follows:
\begin{enumerate*}
\item F associates each object x in $O_1$ with an object Fx in $O_2$;
\item F is a homomorphism from $G_1$ to $G_2$ ;
\item F is a homomorphism from $M_1$ to $M_2$:
\begin{enumerate*}
\item F associates each morphism f in $M_1$(x; y; t) with a morphism Ff in $M_2$(Fx; Fy; Ft),
\item For each x in $O_1$; FIdx = $Id_{Fx}$, and
\item F(f $\circ$ g) = Ff $\circ$ Fg, where ° stands for morphism composition.
\end{enumerate*}
\end{enumerate*}
\end{definition}

\begin{definition}
A pseudo functor F between two pseudo-categories $K_1$ and $K_2$ is the same as Definition 6 with modification: Replace 3(c) of Definition with: Fh = Ff $\circ$ Fg for all f$\in$ M(x, y, t), g$\in$ M(y, z, s) and
h = f $\circ$ g $\in$ M(x; z; t x s).
\end{definition}

For example, if $I_i$ in the pseudo-category BA is linked to $C_j$ in the pseudo-category AA, it means that $C_j$ implement $I_i$. 

\subsection{Problem Definition}
Given the background discussed previously, we are now in a position to define the problem. First we would like to point out that the linkage between business architecture and application architecture will happen at a capability level. We will use the TOGAF definition of business capability: as a synonym for a macro-level business function, \cite{opengroup:togaf}. Business capabilities are services that a business or enterprise offers or requires. A business architecture consists of several capabilities. A capability is realized using an application or a component, both part of an application architecture. We will use a concrete example going forward to define the problem as well as other discussions later in this paper.

\begin{figure}
\centering
\includegraphics[width=3.00in,height=1.886in]{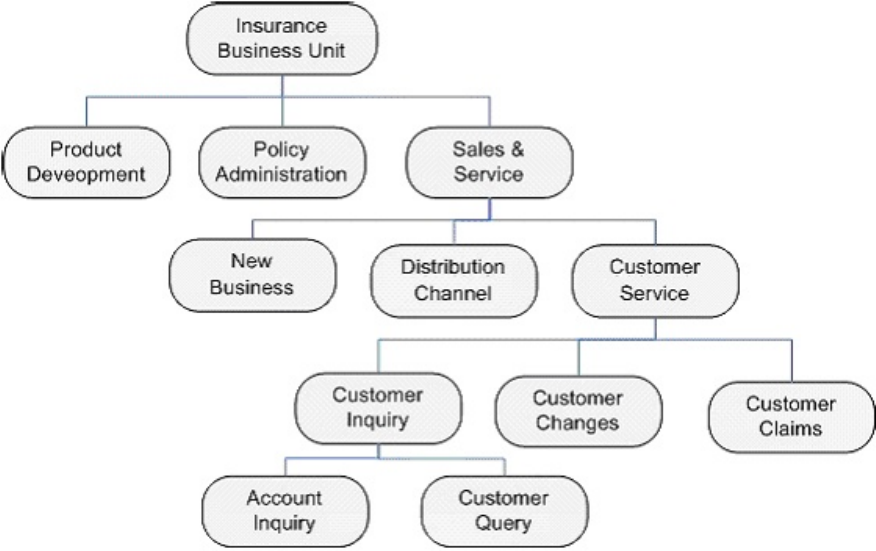}
\caption{Business Capability (partial) - Insurance}
\label{fig1:bcap}
\end{figure}

The problem context that we will address is from the Insurance domain, where some of the business capabilities are as addressed in Figure \ref{fig1:bcap}.

The high level capabilities are Product Development, Policy Administration, Sales and Service etc. Each of these capabilities are further broken down to sub levels. We will focus on the Customer Inquiry capabilities, which currently is composed of Account Inquiry and Customer Query. See  \cite{IGI:978-1-4666-2199-2} for details.

Business usually comes up with new requirements, could be enhancement of an existing capability or adding new capability. The architecture team has several options to address the requirement:

\begin{enumerate*}
\item See if the requested functionality is already existing (in application architecture)
\item See if we can adapt an existing functionality (from component(s) of the application architecture)
\item See if we can assemble the new functionality from existing ones (from component(s) of the application architecture)
\item Create new component
\end{enumerate*}

The problem addressed in this paper is related to items 1 and 2 above. How do we know the request can be met by an existing component(s) or an existing component can be adapted to deliver the functionality.

{\bf Problem Definition:} Given a contract $Ctr_{new}$ that implements an interface $Int_{new}$, (a) find a component \verb+Comp+ that can satisfy the contract, or (b) find a component \verb+Comp+ that can be adapted to implement the functionality. 

The first problem, (a), is referred to in the literature as component matching and second  one, (b), as component adaptation. Our focus in this paper will be on the problem of component matching and the subject od component adaptation will be addressed in furture work. We will, however address the work in this area in the section that follows.
\section{Related Work}\label{sec:related}
\subsection{Component Matching}\label{sec:relatedmatch}
Rollins and Wing \cite{Rollins91specifications} proposed an approach based $\lambda$Prolog to be used as the query as well as specification language. The specification of a component may consist of name, signature and pre- and post-conditions on the behavior of the component. The requirement (query) is also specified using $\lambda$Prolog and then matching betwen the qery and a set of specifications is performed (\verb+satisfy+ as logical implication). They also propose measurements such precision and recall to narrow the search results (closer to the query satisfaction). The main draw back is the performance involved to do the specification matching, when practical size for the set of components is involved.

Zaremiski and Wing \cite{Zaremski:1995:SMT:210134.210179}, \cite{Zaremski:1996:SSM:923008} consider the problem of efficient signature matching as opposed to the specification matching, which also matches the component behavior. The matching procedures are implemented in ML and statistics on the performance when using a variety of matching procedures are provided. Two types of components are considered, function and module. A function is nothing but a function or method declared in a particular software artifcat, for example mathematical functions. The matching is based on the function's type, that is types of both input and output parameters. The module on the other hand is a group of functions - e.g. C++ or Java Class, Ada packages etc. In this case the type is based on the an interface consisting of user defined type and function types. The matching is performed exactly as above but using the interface. The paper provides algorithms for exact match as well as several relaxation matching. 

Zaremski and Wing \cite{Zaremski:1997:SMS:261640.261641} extends their earlier work on signature matching to include the specification matching.  The matching uses Larch/ML interface language to state the pre- and post- conditions and use theorem proving (using Larch) to determine match and mismatch. The matching is extended to function and module as is the case in their previous research. Various "qualities" of matching methods are defined and two application use cases one for for retrieval for reuse and other for subtyping of object-oriented types are presented.

Goguen et. al \cite{DBLP:journals/jsi/GoguenNMLZB96} proposes a component search approach combining the simplicity of key word search for performing faster search with specification-based search for very accurate matching. The concept of ranking is introduced to qualify the query matches. Another idea is the use of multi-level filtering to improve the efficiency of search. The filtering criteria includes semantic filtering. The detailed discussion of these ideas are beyond the scope of this paper and refer the paper cited.

Mili et. al \cite{605762} discusses the design and implementation of storing and retrieving software components from a repository based on formal specifications. The formal specification is called relational specification. The query for retrieval is also a relational specification. The specification is of the form (S,R), where S is a set called space of the specification, nothing but the attributes and variables that will be used in any program on S, and R is a relation on S, a subset of SxS. An ordering between specification is utilized to organize the storage structure of components in the repository which will also allows efficient retrieval using queries. Another interesting aspect of this implementation is that when the system can not find a match, components that come closer to satisfying the query can be retrieved.  

Pahl \cite{DBLP:conf/fm/Pahl01} uses description logic to develop an ontology for matching of components. The description logic consists of three types of entities names objects,concepts and relationships between concepts. The component description and matching ontologies are developed using this. A component is described using the functional behavior and interaction protocol, which together forms the contract for a component. The matching process involves subsumption (relationship involving concepts and roles),  matching of component operation descriptions and matching of component interaction protocols.

There are several other works related to component matching and we provide the references here, Penix and Alexander \cite{DBLP:journals/ase/PenixA99}, Morel and Alexander \cite{10.1109/ASE.2003.1240302} Wang and Krishanan \cite{WangKrishnan06}, Lau et. al \cite{LauNg:12} and related to service matching in web services, Heckel et. al \cite{DBLP:journals/entcs/HeckelCL04}, \cite{DBLP:conf/icalp/EngelsH00}, Iribarne et. al \cite{Iribarne2004} and Iribarne and Vallccillo \cite{IV00}.

\subsection{Component Adaptation}\label{sec:relatedadapt}
Canal et. al \cite{DBLP:journals/Lobjet/CanalMP06} provides detailed definition of software adaptation and provides its characterization in the context of software engineering and its relation to Coordination. They provide some key observations about the need for adaptation- while in the case of hardware components it is possible to select the appropriate component based on it's specification, corresponding selection can not be made in the case of software. The authors states that clearly this is due to the fact that software components are rarely used "as it is" and some kind of adaptation is required. The "mismatch" can arise at different levels - signature (parameters, return values, exceptions), behavior (protocol, service implementation), semantics, and service level.

Kell \cite{Kelljucssurvey} provides a detailed survey about the broad range of relevant research in adaptation and compares the different approaches. The author considers the case of software reuse and calls the two specific difficulties of designing a system in pieces such that each might individually be reused (\emph{modularization}) and composing a required system out pieces readily available for reuse (\emph{mismatch}). Adaptation is defined as the additional work necessary to join the pieces overcoming the mismatch.

Canal et. al \cite{DBLP:journals/tse/CanalPS08} present an approach for software adaptation based on the notions of synchronized vectors and transition systems. The synchronous vector denotes communication between several components using events (keeps track of the event names and components). The labeled transition system (LTS) represents the behavior of the component. The synchronous product of a set of LTS is then defined to capture the dead-lock mismatches, deadlock-freedom is considered as the basis for mismatch. The adaptation contract is then maintained using the synchronous vector mentioned earlier. The paper then presents adaptations for several scenarios such as signature mismatch, behavior mismatch. The paper also discusses the Adaptor tool that was implemented based on the approaches described.

Farias and Sudholt, \cite{farias-sudholt.doa2002} considers various operators for the construction of components which satisfy a correctness property which allows the one component to be substituted by another. The approach is based on the (explicit) protocols that define sequences of possible interactions between request and response between components. A component in this case consists of a set of method declarations, a set of method implementations along with a protocol and corresponding state associated to the protocol. Several protocol operators have been defined and illustrated using the JavaBeans and EJB, two of the widely used component models.

\section{Use Cases for Component Matching}
\label{sec:usecase}
We will discuss two use cases to test the component matching framework discussed in the next section; the first use case identifies a match of an existing application component to meet the business requirement and the other that will require development of a new application component.  \\ 

\noindent{\bfseries Use Case 1: Manage Portfolio} The new business requirements involve providing the customers to group their accounts into one or more portfolios. The business functions to support this involves creating portfolios, deleting portfolios, adding (deleting) accounts to(from) portfolios, transfer accounts between portfolios, see Listing \ref{portfolio}.

\lstdefinestyle{java}{
basicstyle=\small,
language=Java, 
rulecolor=\color{black},
breaklines=true,
numbers=left,
numbersep=5pt, 
numberstyle=\color{black},
commentstyle=\color{black},
keywordstyle=\bfseries\color{black},
stringstyle=\ttfamily
}
\lstset{style=java}
\begin{lstlisting}[caption={Business Specification for Portfolio},label=portfolio]
Interface ManagePortfolio {
 createPortfolio(String portfolioName);
 deletePortfolio(String portfolioName);
 addAccount(Account account, String portfolioName);
 deleteAccount(Account account, String portfolioName);
 transferAccount(Account account, String fromPortfolio, String toPortfolio);
}
\end{lstlisting}

\noindent{\bfseries Use Case 2: Manage Documents} The new business requirements are to create a business component to manage documents by the customers. The business functions to support this involves view documents, search documents, set preferences for a document (e.g. notification when a document is available), see Listing \ref{document}.

\begin{lstlisting} [caption={Business Specification for Manage Document},label=document]
Interface ManageDocuments {
 viewDocument(String documentId);
 searchDocuments(String params);
 setPreference(String documentType, String preference);
}
\end{lstlisting}

\section{Framework for Component Matching}
\label{sec:framework}
We follow the approach based on the work of \cite{report350}, \cite{report327}, and \cite{ICTAC2005}, which is a method for the component specification and programming applications using rCOS. We will first cover some basic definitions that will be used in our framework.

We have already defined contracts (Section \ref{sec:prob}) and we extend this definition to address the dynamic behavior. This is achieved by using the notion of guarded design, see \cite{ICTAC2005}. In this case the $\Phi$ maps each method in the corresponding interface $I$ with a guarded design. The dynamics is defined by the set of failures and divergences (F(Ctr), D(Ctr)) of the contract $Ctr$.

Given two contracts, $Ctr_1$ and $Ctr_2$, we can now define the notion of refinement, see \cite{report350} for details. Essentially the refinement notion states that $Ctr_2$ are either equivalent or  $Ctr_1 \sqsubseteq Ctr_2$, in which case $Ctr_2$ provides the same services as $Ctr_1$ and is not easier to diverge or to deadlock than $Ctr_1$. Next we need the concept of complete contract, which is defined in \cite{comppublication10}, and means that no execution of a trace in the protocol $Prot$ from an initial state will enter a blocking state.

Now we can define the specification of a component C using the complete contract:

\begin{definition}\label{def:prot}
A specification of a component $C$ is a triple S = ($G_{ctr}$,$R_{ctr}$,$C_{S}$), where:
\begin{itemize*}
\item $G_{ctr}$ is a complete contract for the provided interfaces of the component;
\item $R_{ctr}$ is a complete contract for the required interfaces of the component;
\item $C_S$ is a protocol that specifies the interactions with the environment (provided methods) as well as of the required methods. 
\end{itemize*}
The component implements the provided contract using the required contract.
\end{definition} 

For example, Listing \ref{compDM} shows the component (in rCOS) that implements the \verb+MangeDocument+ interface (business).

\begin{lstlisting} [caption={rCOS Component DocumentManager},label=compDM]
component DocumentManager {
  //variables and initialization code
 provided interface ManageDocument {//provided methods
   viewDocument(String documentId) {//implementation code;}
   searchDocuments(String params) {//implementation code;}
   setPreference(String documentType, String preference) {//implementation code;}
 }
 internal interface I {//private methods
  connect() {//code to connect to the repositroy; }
 }
 required interface J {//required services
   getDocument(String documentId);
   getDocuments(String criteria);
   updateDocumentSetting(String documentType);  
 }
}
\end{lstlisting}
 
The provided protocol can be expressed using regular expression as:
\begin{verbatim}
(?searchDocument+?setPreference)*|
  (?searchDocument+
    ?viewDocument?setPreference)*
\end{verbatim}
In order to perform matching operation and to determine whether a component can meet the requirements, we can deal with a publication rather than a specification. The publication is derived by removing the guards in the method specifications. We can define the publication contract $Ctr_P$ as:
\begin{definition}
A publication contract $Ctr_P$ is (I,D,T) where:
\begin{itemize*}
\item I is the interface I (provided methods);
\item D is a function that defines each method m of I with a design (no guard);
\item T  is a protocol (set of traces) over the methods m of I.
\end{itemize*}
\end{definition}
Given the above we can define the publication for a component:
\begin{definition}
A publication of a component $C_P$ is (P,R,$C$), where:
\begin{itemize*}
\item P is a publication contract of the interface I (provided methods);
\item R is a publication contract of the interface J such that the methods in J are at least the methods required by the component to implement I;
\item $C$  is a causal relation over the methods in I and J
\end{itemize*}
The causal relation $C$ is defined such that it satisfies the condition: if remove the trace from $C$ that are not it Meth.I, we get the protocol of P, and the same way if we remove  the trace from $C$ that are not it Meth.J, we get the protocol of R.
\end{definition}

The publication of the component described in Listing \ref{compDM} is shown below in Listing \ref{compPub}
\begin{lstlisting} [caption={rCOS Publication of DocumentManager},label=compPub]
publication DocumentManager {
  //variables and initialization code
 provided interface ManageDocument {//provided methods
   viewDocument(String documentId) {//unguarded design;}
   searchDocuments(String params) {//unguarded design;}
   setPreference(String documentType, String preference) {//unguarded design;}
 }
 internal interface I {//private methods
  connect() {//unguarded design; }
 }
 required interface J {//required services
   getDocument(String documentId);
   getDocuments(String criteria);
   updateDocumentSetting(String documentType);  
 }

 provided protocol {
  (?searchDocument+?setPreference)*|  (?searchDocument+ ?viewDocument?setPreference)*
 }

 required protocol {
  //similar to the provided protocol but for the required services.
 }
}
\end{lstlisting}
\subsection{Protocol and Matching}\label{sec:matchC}
The protocol in the publication can be used to determine the interaction of a component and the environment, see \cite{comppublication10}. Let us say, we have the publications $C_{P1}$ and $C_{P2}$, for two components $C_1$ and $C_2$ respectively, and that $C_1$ is an existing component. If the provided protocol of $C_{P1}$ contains all the invocation trace of the required protocol of $C_{P2}$, then we can plugin the component $C_1$ to provide the services that $C_2$ requires. \emph{This will be the basis for our matching process}.

As we are looking to automate the matching process, we need to find a way to generate the protocols for the existing components (or contracts, or publications). We provide some of the key references to this topic on protocol generation: \cite{DBLP:conf/cbse/PochP09}, \cite{report446}, \cite{1132}, \cite{donginterface}, \cite{ZimmermannS06}, \cite{aboth10}, \cite{BothRichter10}, \cite{703524}, \cite{698725}, \cite{789513}, and \cite{720980}with a brief review. 

In \cite{DBLP:conf/cbse/PochP09}, Poch and Plasil presents technqiue for extracting the behavior of component given as a set of Java classes. The behavior is extracted as Behavior Protocol, a high level specification ocapturing the finite sequences of method calls allowed on the component provided and required interfaces, represented as regular expressions. A prototype tool JAbstractor is also discussed that implements the extraction of behavior.

In a series of papers, \cite{report446}, \cite{1132}, \cite{donginterface}, Dong, Liu et.al. present an automata-based interface model to describe the interaction behavior of software components. An input deterministic automata, that define all the non-blockable traces of invocation to services provided by the component. An algorithm is also presented that, for a given component automaton, generates the interface model that has the same behavior as the original automaton. Tool support is not available at this time.

In \cite{BothRichter10} Both and Richter presents a tool based automatic generation of the component protocol to ensure safe behavior of component interactions, based on earlier work, \cite{aboth10}, \cite{ZimmermannS06}. This process is done in five steps that involves the conversion of code to reachability models using symbolic pushdown systems, solving the using the tool HalVRE (Halle's exact Value Range Extractor), and finally constructing the component protocol automaton.

The work in \cite{703524}, \cite{698725}, provide the specification language using the component interaction automata to model all interesting aspects of the interaction of components along with a tool based verification of these specifications, the DiVinE model checking tool.

\subsection{Proposed Approach}
The proposed framework will be developed based on the follwing lines:
\begin{itemize}
\item We have the collection C of components, $C_i$, with $1\leq i \leq N$, in the application architecture AA, see Definition \ref{def:aa}.
\item We generate the protocol for each of these components, see Definition \ref{def:prot}. \emph{This is a one time activity whenever the architecture is defined or modified}.
\item For the new requirement, we generate the required protocol based on one of the algorithms discussed above.
\item We try to find a protocol match by comparing the required protocol with the component protocols in AA. This will be a two-level matching process, with level 1 being the signature matching process discussed in Zaremski and Wing \cite{Zaremski:1995:SMT:210134.210179}, \cite{Zaremski:1996:SSM:923008}, Goguen et. al \cite{DBLP:journals/jsi/GoguenNMLZB96}, followed by the protocol matching described in \cite{donginterface}. If a match is found, then we can "use" this component to provide the services required. If a match is not found, then we can either refine an existing component (found by signature matching) to provide the services or develop a new component.
\end{itemize}
Currently work is in progress to define efficient algorithm to implement the approach defined above.
\section{Concluding Remarks} \label{sec:conclusion}
In this paper the problem of matching business architecture/requirements with the corresponding application architecture is considered.  A framework is proposed that we hope to be the basis for tool support that will be developed subsequently as our research progresses. The goa is to provide an efficient algorithm for the protocol generation and matching. Further work will involve when a match is not available, adapting the existing functionality of a component to meet the requirements.





\bibliographystyle{elsarticle-num-names}
\bibliography{capability}







\end{document}